# LaPtSb: a half-Heusler compound with high thermoelectric performance

Q. Y. Xue, H. J. Liu[*], D. D. Fan, L. Cheng, B. Y. Zhao, J. Shi

*Key Laboratory of Artificial Micro- and Nano-Structures of Ministry of Education and School of Physics and Technology, Wuhan University, Wuhan 430072, China*

The electronic and transport properties of the half-Heusler compound LaPtSb are investigated by performing first-principles calculations combined with semi-classical Boltzmann theory and deformation potential theory. Compared with many typical half-Heusler compounds, the LaPtSb exhibits obviously larger power factor at room temperature, especially for the *n*-type system. Together with the very low lattice thermal conductivity, the thermoelectric figure of merit (*ZT*) of LaPtSb can be optimized to a record high value of 2.2 by fine tuning the carrier concentration.

**1. Introduction**

Owing to serious energy crisis and environment pollution, the demand for sustainable and clean energy source becomes more and more important. Thermoelectric materials can directly convert waste heat into electricity, which have attracted intensive attention from the science and industry communities. The performance of thermoelectric materials is usually determined by the dimensionless figure of merit

$$ZT = \frac{S^2 \sigma}{\kappa_l + \kappa_e} T , \qquad (1)$$

where $S$, $\sigma$, $T$, $\kappa_l$ and $\kappa_e$ are the Seebeck coefficient, the electrical conductivity, the absolute temperature, the lattice and electronic thermal conductivity, respectively. An ideal thermoelectric material behaves as phonon-glass electron-crystal (PGEC), which requires maximize the power factor ($S^2\sigma$) and/or minimize the thermal conductivity ($\kappa_l + \kappa_e$). However, these transport coefficients ($S$, $\sigma$, $\kappa_e$) are usually coupled with each other, and it is still a great challenge to obtain a high $ZT$ value, especially for the bulk materials.

---

[*] Author to whom correspondence should be addressed. Electronic mail: phlhj@whu.edu.cn.

Ternary half-Heusler (HH) system is a class of intermetallic compounds, some of which are potential thermoelectric materials because of their narrow band gaps, high temperature stability, large Seebeck coefficients, and moderate electrical conductivities [1-5]. However, most HH compounds were found to exhibit higher thermal conductivity in the order of magnitude of 10 W/mK, which is undesirable for high performance thermoelectric materials [6]. Over the past two decades, the thermoelectric properties of a few HH compounds have been extensively investigated, especially for the (Ti, Zr, and Hf)NiSn-based and (Ti, Zr, and Hf)CoSb-based compounds [1-17]. Among them, Chen *et al.* reported that a $ZT$ of ~1.2 can be achieved around 900 K for *n*-type $Hf_{0.6}Zr_{0.4}NiSn_{0.995}Sb_{0.005}$ alloys without the need of nanostructures [16]. They further found that a $ZT$ of ~1.3 can be achieved for *n*-type $Hf_{0.65}Zr_{0.25}Ti_{0.15}NiSn_{0.995}Sb_{0.005}$/nano-$ZrO_2$ near 850 K [17]. Fu *et al.* found that the *p*-type $FeNb_{0.88}Hf_{0.12}Sb$ and $FeNb_{0.86}Hf_{0.14}Sb$ alloys exhibit a $ZT$ of about 1.5 at 1200 K [15]. However, these $ZT$ values of HH compounds remain low compared with the target value of about 3.0, which can compete with the efficiency of conventional refrigerators or power generators. To achieve higher $ZT$ value, an efficient strategy is to search thermoelectric materials with intrinsically low thermal conductivities. Recently, Carrete *et al.* [18] theoretically investigated the thermal conductivities of a large number of HH compounds and found that the compound LaPtSb has a particularly low lattice thermal conductivity of 1.72 W/mK at room temperature. It is therefore interesting to ask whether such compound can exhibit better thermoelectric performance. In this work, the thermoelectric properties of LaPtSb are predicted by multiscale calculations which combine the first-principles calculations, Boltzmann theory, and deformation potential theory. We demonstrate that by fine tuning the carrier concentration of LaPtSb, a maximal $ZT$ value of 2.2 can be achieved for *n*-type system at room temperature, which makes it a very promising candidate for high performance thermoelectric materials.

## 2. Computational Methods

The structure optimization and electronic structure calculations of LaPtSb are performed by using the projector-augmented wave (PAW) method [19], as implemented in the Vienna *ab initio* simulation package (VASP) [20, 21, 22]. The treatment of exchange and correlation energy is adopted in the form of Perdew-Burke-Ernzerhof (PBE) [23] with the generalized gradient approximation (GGA). It's necessary to add a Hubbard-type term $U$ [24] to act on the La 5$d$ and Pt 5$d$ states, which can greatly improve the quality of PBE and give the correct electronic structure [25]. We use an effective Hubbard $U$ of 8.1 eV for La and 3.0 eV for Pt, as obtained from the AFLOWLIB.org consortium repository [26]. The plane wave cut-off energy is set to be 400 eV, and the Brillouin zone is sampled with a Monkhorst-Pack $k$-mesh of 15×15×15. Optimal atomic positions are determined until the magnitude of the forces acting on every atom becomes less than 0.01 eV/Å.

The electronic transport coefficients are evaluated by using the semi-classical Boltzmann transport theory [27] with relaxation time approximation. In terms of the so-called transport distribution function $\Xi(\varepsilon) = \sum_{\vec{k}} \vec{v}_{\vec{k}} \vec{v}_{\vec{k}} \tau_{\vec{k}}$, the Seebeck coefficient $S$ and the electrical conductivity $\sigma$ are given by

$$S = \frac{ek_B}{\sigma} \int d\varepsilon \left( -\frac{\partial f_0}{\partial \varepsilon} \right) \Xi(\varepsilon) \frac{\varepsilon - \mu}{k_B T}, \qquad (2)$$

$$\sigma = e^2 \int d\varepsilon \left( -\frac{\partial f_0}{\partial \varepsilon} \right) \Xi(\varepsilon), \qquad (3)$$

Here $f_0$ is the equilibrium Fermi-Dirac distribution function, $\mu$ is the chemical potential, $k_B$ is the Boltzmann's constant, and $\vec{v}_{\vec{k}}$ is the group velocity at state $\vec{k}$. The calculations of transport coefficients are performed in the so-called BoltzTraP code [27]. The relaxation time $\tau_{\vec{k}}$ is estimated by the deformation potential (DP) theory as proposed by Bardeen and Shockley [28]. In addition, the electronic thermal conductivity $\kappa_e$ is calculated via the Wiedemann-Franz law [29]

$$\kappa_e = L\sigma T, \tag{4}$$

Here the Lorenz number $L$ can be obtained by [30, 31]

$$L = \frac{k_e}{\sigma T} = \left(\frac{k_B}{e}\right)^2 \left[\frac{3F_2}{F_0} - \left(\frac{2F_1}{F_0}\right)^2\right], \tag{5}$$

where $F_i$ is a function of the reduced Fermi energy $\varsigma$ given by $F_i(\varsigma) = \int_0^\infty \frac{x^i dx}{e^{(x-\varsigma)}+1}$.

## 3. Results and Discussions

HH compounds crystallize into the cubic MgAgAs-type structure [32] with the space group $F\bar{4}3m$ [33]. Figure 1 shows the unit cell of the HH compound LaPtSb, where La atoms are located at the position of 4$a$(0,0,0), Pt atoms at the 4$c$(1/4,1/4,1/4), and Sb atoms at the 4$b$(1/2,1/2,1/2) in Wyckoff coordinates. The structure can also be regarded as four interpenetrating *fcc* lattices, containing a lattice of La atoms, a lattice of Pt atoms, a lattice of Sb atoms, and a lattice of vacancies. The La and Sb atoms form a rock salt structure. The optimized lattice constant of LaPtSb is 6.87 Å. The total valence electron count (VEC) in the primitive cell satisfies the criterion of VEC=18 [34], as generally found in HH compounds.

We first focus on the electronic band structure of LaPtSb, which is shown in Figure 2 along the high-symmetry lines of the irreducible Brillouin zone (IBZ). We can see that both the valence band maximum (VBM) and the conduction band minimum (CBM) are located at Γ point, resulting in a direct band gap of 0.23 eV. Moreover, we find that the energy band near CBM is rather sharp, which indicates a quite small effective mass. In contrast, the energy band near VBM is relatively flat, and has a threefold degeneracy. All these observations will play a significant role in the transport properties of LaPtSb, as discussed in the following.

To evaluate the electronic transport coefficients within the framework of Boltzmann theory, the relaxation time has to be determined which is usually very complicated since it depends on the detailed scattering mechanisms. As the wave length of thermally activated carriers at room temperature is much larger compared with the lattice constant, the scattering of carriers is dominated by the electron-acoustic phonon coupling [35]. In this respect, we evaluate the relaxation time with the help of deformation potential theory, which can effectively describe the electron-acoustic phonon interactions. For three-dimensional system, the relaxation time along the $\beta$ direction at temperature $T$ can be expressed as:

$$\tau_\beta = \frac{2\sqrt{2\pi} C_\beta \hbar^4}{3(k_B T m_{dos}^*)^{3/2} E_\beta^2}, \tag{6}$$

In this formula, $C_\beta$ is the elastic constant defined as $C_\beta = \frac{1}{V_0} \frac{\partial^2 E}{\partial (\Delta l / l_0)^2}\bigg|_{l=l_0}$, where $E$ is the total energy of the system, $l_0$ is the lattice constant along the direction of $\beta$, $\Delta l = l - l_0$ is the corresponding lattice distortion, and $V_0$ is the equilibrium volume of the unit cell. The deformation potential constant $E_\beta$ is calculated as $E_\beta = \frac{\partial E_{edge}}{\partial (\Delta l)/l_0}$, which represents the shift of band edges (VBM or CBM) per unit strain. The density-of-states effective mass $m_{dos}^*$ is given by $m_{dos}^* = \left(m_x^* m_y^* m_z^*\right)^{\frac{1}{3}}$, where $m_x^*$, $m_y^*$, and $m_z^*$ are the effective mass along the $x$, $y$, and $z$ directions, respectively. These three quantities ($C_\beta, E_\beta, m_{dos}^*$) can be readily obtained from first-principles calculations. It should be noted that we must consider the band degeneracy mentioned above when calculating the $m_{dos}^*$ [36]. The derived relaxation time of LaPtSb compound is summarized in Table I for both *p*- and *n*-type carriers. It is obvious to find that the relaxation time of electrons is much larger than that of holes, which mainly originates from a very small density-of-states effective mass of the former, as can be also seen from the dispersion relations around CBM and VBM in Fig. 2. Note that such significant difference in the relaxation time between electrons and holes has been also found in other systems such as few-layer black phosphorus and phosphorene [37, 38]. On the other hand, the relaxation time of both electrons and holes are larger than those of typical thermoelectric materials such as $Bi_2Te_3$ [39] and $MoS_2$ [35], which suggests the favorable thermoelectric performance of LaPtSb compound.

In Figure 3, we plot the calculated Seebeck coefficient $S$, the electrical conductivity $\sigma$, and the power factor $S^2\sigma$ of LaPtSb as a function of carrier concentration at room temperature. We can see that the absolute values of Seebeck coefficient decreases with increasing carrier concentration and becomes vanishing when the concentration is larger than $10^{21}$ cm$^{-3}$ for both *p*-type and *n*-type systems. On the other hand, we find that the electrical conductivity of *n*-type system is much larger than that of *p*-type system at the same carrier concentration, which is caused by the obvious difference between the relaxation time of electrons and holes. Comparing Fig. 3(a) with 3(b), we note there is a sharp increase of the electrical conductivity at the carrier concentration where the Seebeck coefficient has very small absolute value. It is thus natural to find a balance between them so that the power factor can be maximized by fine tuning the carrier concentration. Indeed, we see from Fig. 3(c) that the power factor of LaPtSb can reach peak values at particular concentration for both electrons and hole. For example, at a concentration of $2.7\times10^{20}$ cm$^{-3}$, the power factor of *n*-type system can be optimized to as high as 0.43 W/mK$^2$, which is much larger than those of many traditional thermoelectric materials such as $Bi_2Te_3$ [39]. Such characteristic suggests again that the HH compound LaPtSb could have favorable thermoelectric performance.

To evaluate the *ZT* value of LaPtSb compound, one also needs to know the thermal conductivity, which includes the contributions from both phonons ($\kappa_l$) and charge carriers ($\kappa_e$). Using the Wiedemann-Franz Law given by Equation (4), the electronic thermal conductivity can be readily obtained and is shown in Fig. 3(d) as a function of carrier concentration at room temperature. We see that the behavior of $\kappa_e$ is basically the same as that of electrical conductivity shown in Fig. 3(b). This is reasonable since the Lorenz number calculated from Equation (5) has little change with the carrier concentration. For the phonon part, we take the result of Carrete *et al.* where a room temperature thermal conductivity of 1.72 W/mK is predicted by solving phonon Boltzmann transport equation [18]. Such a low value of thermal conductivity is very desirable for thermoelectric applications when compared with other HH compounds.

We can now evaluate the thermoelectric performance of LaPtSb compound by inserting all the transport coefficients into Equation (1). Figure 4 plots the room temperature $ZT$ values as a function of carrier concentration. For the *p*-type system, a maximal $ZT$ of 1.3 appears at the concentration of $1.2 \times 10^{19}$ cm$^{-3}$. The $ZT$ value can be further enhanced to 2.2 for the *n*-type system when the carrier concentration is tune to $3.3 \times 10^{19}$ cm$^{-3}$. It should be mentioned that such a $ZT$ value exceeds those of many good thermoelectric materials, and in particular the typical (Ti, Zr, and Hf)NiSn-based and (Ti, Zr, and Hf)CoSb-based HH compounds. In the case of rare-earth based HH compounds, although there is currently no experimental study on LaPtSb, we note that the thermoelectric properties of several other rare-earth based HH systems, such as LnPdSb (Ln=Ho, Er, Dy), ScMSb (M=Ni, Pd, Pt), YMSb (M=Ni, Pd, Pt), ErPdX (X=Sb, Bi), PtYSb, and YNiBi have been experimentally reported [40-45]. Compared with the LaPtSb predicted in the present work, these HH compounds exhibit relatively smaller power factors and larger lattice thermal conductivities. As a result, the thermoelectric performances are less favorable than that of LaPtSb.

## 4. Summary

In summary, our theoretical work demonstrates that the HH compound LaPtSb could be optimized to exhibit very good thermoelectric performance, which is believed to be benefited from the relatively large power factor and low lattice thermal conductivity. At a moderate carrier concentration of $3.3 \times 10^{19}$ cm$^{-3}$, the room temperature $ZT$ value can be enhanced to as high as 2.2 for *n*-type system, which exceeds that of the HH compounds reported so far. Our calculations suggest that the *n*-type LaPtSb is a very promising thermoelectric material and major efforts should be directed to developing comparatively effective *p*-type LaPtSb so that efficient thermoelectric modules based on this HH compound could be realized.

**Acknowledgements**

This work was supported by the National Natural Science Foundation (Grant No. 11574236 and 51172167) and the "973 Program" of China (Grant No. 2013CB632502).

**Table I** The deformation potential constant $E_1$, the elastic constant $C$, the density-of-states effective mass $m_{dos}^*$, and the relaxation time $\tau$ of LaPtSb at room temperature.

| Carrier type | $E_1$ (eV) | $C$ | $m_{dos}^*$ ($m_e$) | $\tau$ (fs) |
|---|---|---|---|---|
| electron | 21.73 | 1.45 | 0.03 | 2063 |
| Hole | 14.43 | 1.45 | 0.25 | 194 |

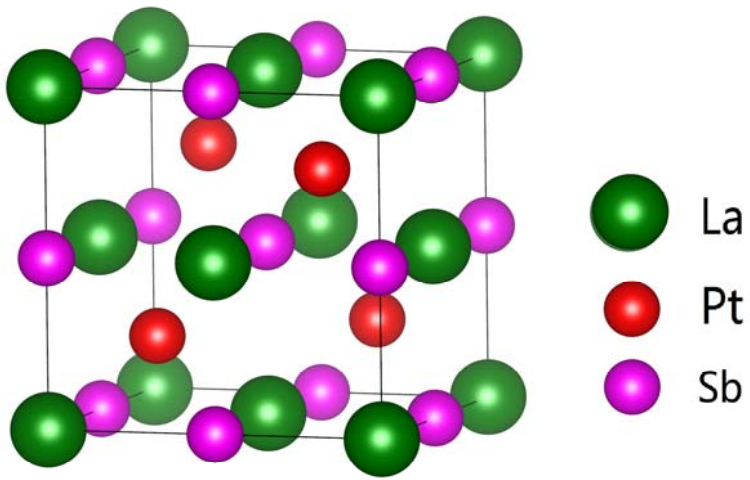

**Figure 1** (Color online) The unit cell of the half-Heusler compound LaPtSb.

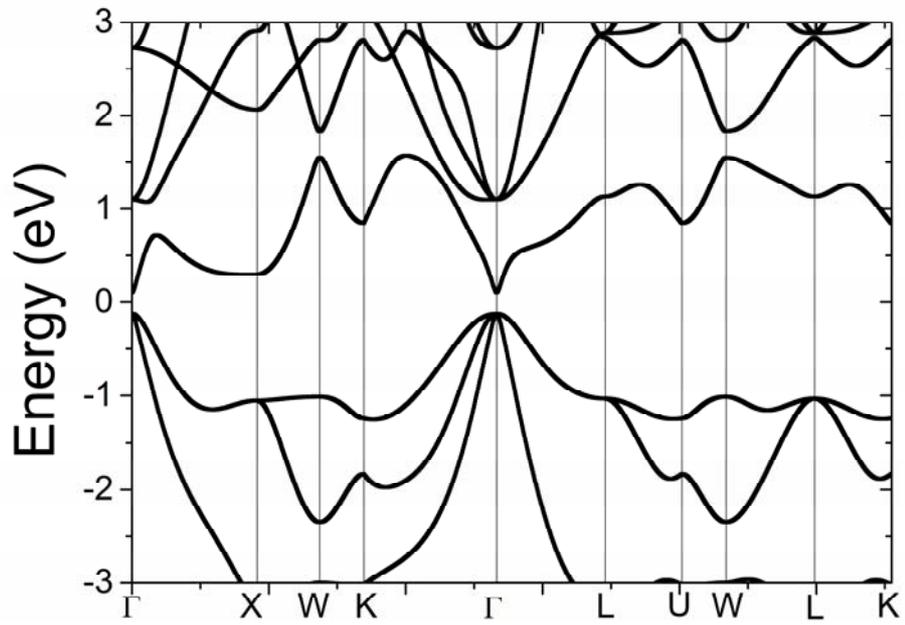

**Figure 2** The calculated energy band structure of LaPtSb.

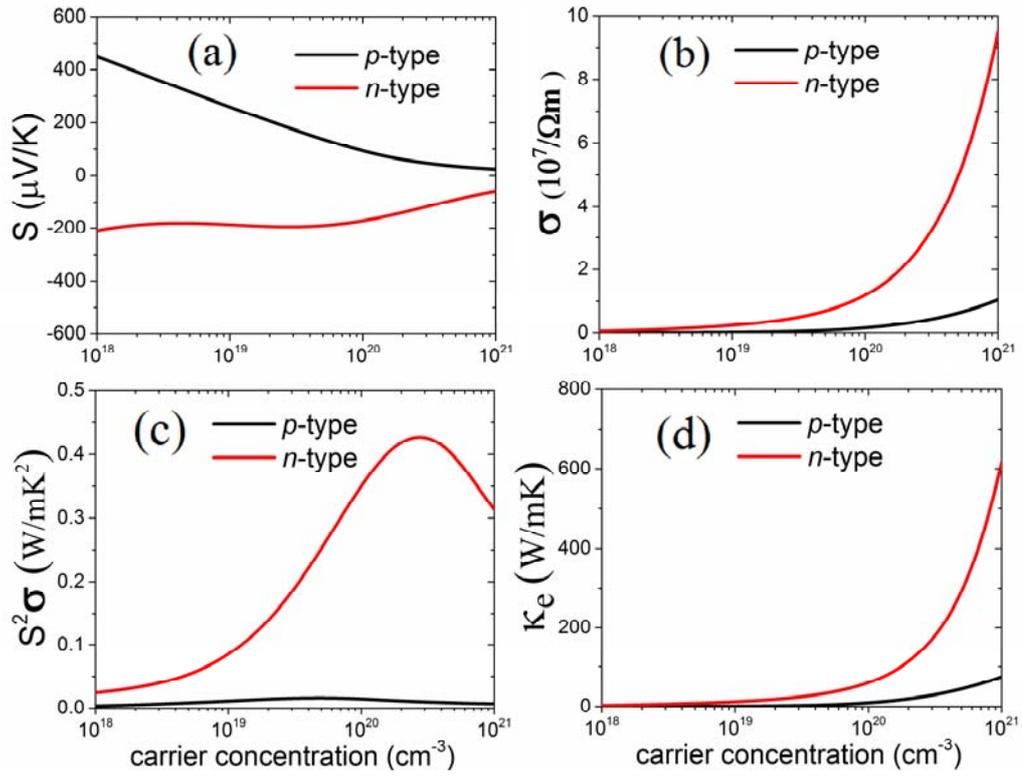

**Figure 3** The calculated electronic transport coefficients of LaPtSb: (a) the Seebeck coefficient $S$, (b) the electrical conductivity $\sigma$, (c) the power factor $S^2\sigma$, and (d) the electronic thermal conductivity $\kappa_e$ as a function of carrier concentration at room temperature. Negative and positive carrier concentrations represent *n*- and *p*- type carriers, respectively.

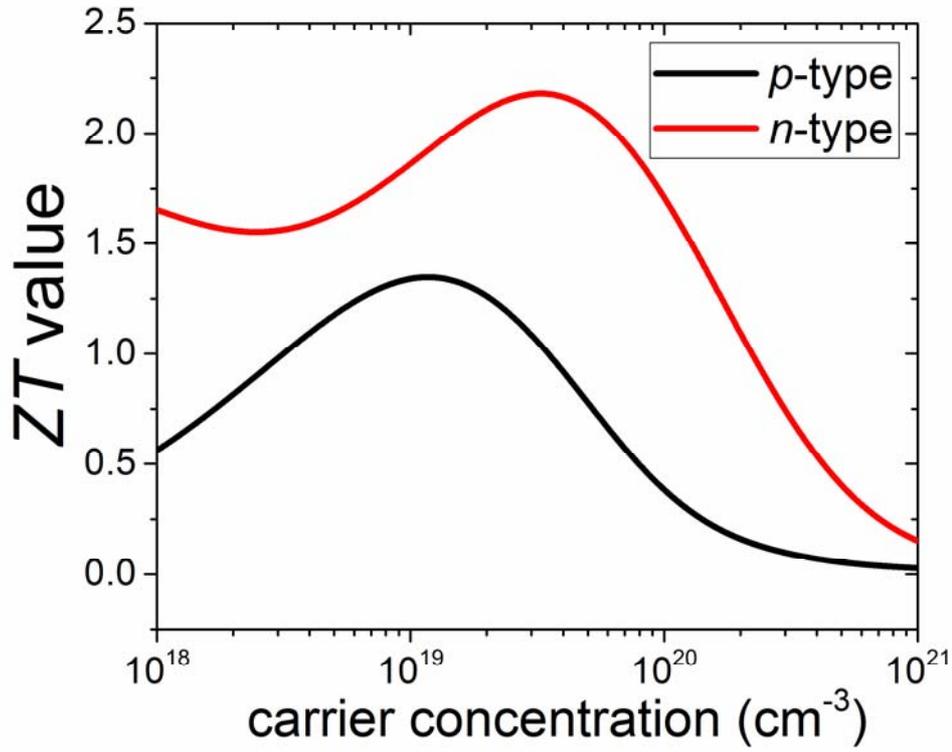

**Figure 4** The calculated $ZT$ values of LaPtSb as a function of carrier concentration at room temperature.